# Correlated X–ray timing and spectral behavior of GX 349+2: RXTE PCA data

Vivek K. Agrawal[1] and Sudip Bhattacharyya[2,3]

[1] ISRO Satellite Centre, Airport Road, Bangalore 560 017, India

[2] Joint Astronomy Program, Indian Institute of Science, Bangalore 560012, India

[3] Indian Institute of Astrophysics, Bangalore 560 034, India
(vivek_isac@yahoo.com; sudip@physics.iisc.ernet.in; sudip@iiap.ernet.in)

**Abstract.** We present a detailed and systematic investigation of correlated spectral and timing properties of the Z source GX 349+2, using a huge amount ($\sim$ 40 ks) of good quality data from September 29, 1998 to October 13, 1998, obtained with the Proportional-Counter-Array on-board RXTE satellite. We, for the first time, give a detailed comparison between the normal-branch-properties and the flaring-branch-properties of this source, as it shows an extended normal branch (which is a rare phenomenon for it) for our case. In our work, the properties of the peaked noise are analyzed as functions of the position on the Z-track (in color-color-diagram), and they are discussed in connection with the peaked noise seen for another Z source Cyg X-2 (at low overall intensities). This will help to construct theoretical models for the peaked noise, as well as to understand the physics behind the Z-shaped tracks traced by these kind of sources in color-color-diagram. We also find a QPO (centroid frequency $\sim$ 3.8 Hz) at the upper part of the flaring branch. This is for the first time, a QPO is seen from this source and hence is very important in understanding the nature of its X–ray emitting components.

**Key words.** accretion, accretion disks — stars: binaries: close — stars: individual (GX 349+2) — stars: neutron — X–rays: stars

## 1. Introduction

The bright low-mass X-ray binary (LMXB) GX 349+2 (also called Sco X-2) belongs to a class, called Z sources (Hasinger and van der Klis 1989). These are the most luminous

X–ray binaries, which are believed to contain neutron stars as the accreting objects, since two of them viz, Cyg X-2 (Smale 1998) and GX 17+2 (Kuulkers et al. 2001) have exhibited type I X-ray bursts, which are the characteristic of the neutron star. A Z source traces out a Z-shaped track in X–ray color-color diagram (CD) and hardness-intensity diagram (HID). The Z-track generally consists of three parts, horizontal branch (HB), normal branch (NB) and flaring branch (FB). It is generally believed that the inferred-mass-accretion-rate increases along the Z-track from HB to FB (Hasinger et al. 1990). So far six Z sources have been discovered, which are further divided into two subclasses: (1) Cyg-like: Cyg X-2, GX 5-1 & GX 340+0, and (2) Sco-like: Sco X-1, GX 349+2 & GX 17+2 (Kuulkers et al. 1994, 1997). Sco-like objects have smaller and slanted HBs and larger FBs, while HBs for Cyg-like sources are comparatively larger and horizontal, and their FBs are generally much smaller. The long term variation in shape and position of Z-track (secular motion) has been observed for Cyg-like source (Kuulkers et al. 1994, 1996; Kuulkers and van der Klis 1996). It has been suggested that Cyg-like sources are being viewed at higher inclination angle compared to Sco-like sources (Kuulkers et al. 1994) and contain neutron stars of higher magnetic field strength (Psaltis et al. 1995). Quasi-periodic-oscillations (QPO) and noise components are found in the power spectra of Z sources. These features are generally well-correlated with the position of the source on the Z-track (Hasinger and van der Klis 1989; van der Klis 1995). There are three types of common noise: very-low-frequency-noise (VLFN), low-frequency-noise (LFN) and high-frequency-noise (HFN). The noise components VLFN and HFN are common in all the three branches, but LFN is observed only in HB. QPOs with frequencies in the range $15-60$ Hz are generally observed in HB and in the upper parts of NB (van der Klis 1995). These are called horizontal-branch-oscillations (HBO). A QPO with the frequency in the range $5-8$ Hz is observed from the middle part of NB up to the NB/FB vertex and is called normal-branch-oscillation (NBO). A sudden increase in centroid frequency of NBO is observed at the NB/FB vertex and according to the common belief, NBO transforms to FBO (flaring-branch-oscillation) at this point (Dieters and van der Klis 2000). As the source moves up along FB, both centroid frequency and full-width-half-maximum (FWHM) of FBO increases. In addition to the low frequency QPOs, kHz QPOs ($200-1200$ Hz) are also observed for all the Z sources (see van der Klis 2000, for a review).

GX 349+2 is very similar to Sco X-1 in many respects. For example, both objects exhibit strong flaring behavior and the orbital periods are also similar ($P_{\rm orb} \sim 18.9$ hr for Sco X-1 and $\sim 22$ hr for GX 349+2). However, some properties of GX 349+2 are different from those observed for the other five Z sources. First, it never showed a horizontal branch. Besides, instead of NBO and FBO (together called N/FBO), a broad peaked noise with a centroid frequency and FWHM of around 6 Hz and 10 Hz respectively were observed in its FB (EXOSAT observation; Ponman et al. 1988). It was found that the width of the peaked noise component decreases with increasing intensity, which is contrary to the suggestion made by several models (Lamb et al. 1985; Boyle et al. 1986; Hameury et al. 1985). Ponman et al. (1988) noticed that the strength of the peaked noise is maximum in the intermediate intensity band, i.e., at the NB/FB vertex. However in their work, the lower part of FB and NB could not be differentiated clearly, since they divided the data according to the intensity and not according to the position along the Z curve. They also investigated energy dependence of peaked noise properties and found that rms strength of peaked noise was higher in the higher energy bands and there was no time lag between hard and soft photons. Observation with GINGA Large-Area-Counter indicates that the width and the centroid frequency of peaked noise component does not vary significantly as the source moves along FB (O'Neill et al. 2001). They found that the strength of the peaked noise are maximum in the lower part of FB ($\sim 10\%$ of the way up the FB) and it becomes weaker as the source moves up the FB. Kuulkers & van der Klis (1998) reported the detection of a similar peaked noise component at the lower part of FB using $\sim 4$ hrs of RXTE observation. A broader and somewhat weaker peaked noise was detected when the source was in the NB. They also showed that rms strength of peaked noise increases with increasing photon energy. Till now for GX 349+2, a narrow NBO or FBO has never been observed.

Inspite of being unique (Kuulkers & van der Klis 1998) among the Z sources, GX 349+2 is the least-understood one, and even not well-observed. In this paper, for the first time we present a detailed quantitative study of this source with good quality RXTE PCA data. A similar peaked noise (as observed in GX 349+2) was observed for the Cyg-like Z source Cyg X-2 at low overall intensities (Kuulkers et al. 1999). We therefore compare the Z curves for these two sources, in order to understand the physics behind the peaked noise. In the absence of a proper model, our detailed study is expected to help formulate one.

In section 2, we describe the observations and the procedure of analysis. We summarize the results in section 3 and discuss the implications in section 4.

## 2. Observations and Analysis

We analyze RXTE-PCA public archival data from September 29 to October 13, 1998. The other observations of this year (in January, 1998) were analyzed by Zhang et al. (1998) and they reported the finding of twin kHz QPOs. We use the data for which all 5 PCUs were on. The total good amount of data available for our analysis is 39.8 ks (details of observations are shown in Table 1). For the data reduction and analysis, we use standard FTOOLS package version 5.0 distributed and maintained by NASA. The Standard 2 mode data with time resolution of 16 s and effective energy range of 2-60 keV are used for the spectral analysis. The background subtraction is applied on the data before creating the X-ray spectra. The sky_VLE model of epoch 3 for bright source is used to calculate the PCA background. X-ray color-color diagram and hardness-intensity diagram are constructed using 256 s averages. We define the soft color as the count-rate-ratio between 3.5-6.4 keV and 2-3.5 keV energy bands and hard color as the count-rate-ratio between 9.7-16 and 6.4-9.7 keV energy bands. The intensity is defined as the count rate in the energy band 2-16 keV. To define the position of the source along the Z-track, the 'rank number' or '$S_z$' parameterization technique was first introduced by Hasinger et al. (1990) and has been further modified (Hertz et al. 1992; Kuulkers et al. 1994; Dieters and van der Klis 2000). We use the current version of this technique modified by Dieters and van der Klis (2000). We select the normal points in the CD such that they form a smooth curve The color-color points in the CD are projected on this smooth curve. The $S_z$ parameter for each projected points were calculated by measuring their distance from NB/FB vertex. We choose two reference points: NB/FB vertex ($S_z = 2$) and the end point of the FB ($S_z = 3$). The rest of the Z-track is normalized using the length of FB. The power spectra for all observations are created using science event mode (E_8$\mu$s_32B_14_1s) data of 8$\mu$s time resolution and energy range 5.1-60 keV. We create power spectra for 8 s intervals and then averaged over the time intervals that correspond to the change in the mean $S_z$ parameter by $\leq$ 10%. The average power spectra are rebinned afterwards. The power spectra are fitted by a simple power law (between 0.125-150 Hz; representing the VLFN) and a Lorentzian (denoting the peaked noise). To study the energy dependence of noise components, we create power spectra in the energy bands, 2.0-5.1, 5.1-7.0, 7.0-10.0, 10.0-16.0 keV. We use SB_250_250$\mu$s_0_13_2s mode data of 244$\mu$s time resolution to extract the power spectra in the energy band 2.0-5.1 keV. Errors on the fitting parameters are calculated by using $\Delta\chi^2$=1.0 (68% confidence).

## 3. Results

The complete X-ray color-color diagram for 14 days observations consists of an extended FB and comparatively smaller and vertical NB (Figure 1). The solid curve in Figure 1 represents an approximate Z-track passing through the normal points. The hardness-intensity diagram for these observations is shown in Figure 2. The intensity of the source increases by a factor of two as it moves from NB/FB vertex to the top of FB, but it does not vary much along NB, since the NB is almost vertical. It is noticed that FB is more scattered at higher intensities. We find that during our observations, the source occupies different parts and covers different fractions of the Z-track (Table 1). During the most of the observations, source is found in various parts of FB and it spends $\sim$ 79% of total good time in this branch. The variation of mean $S_z$ values (from Table 2) on time scale of several days is shown in Figure 3. There are only a few points below $S_z = 2$ and therefore, the occurrence of the NB is a rare phenomenon. However, unavailability of continuous data does not allow us to say about the motion of the source between two consecutive $S_z$ values in this figure. We find that during observation no. 7, source travels from lower part of FB to upper part of FB in a time scale of $\sim 30$ minutes and covers maximum fraction of Z-track ($\sim 68\%$). The transition from NB to FB or vice versa is detected during three observations, and out of these three, the source traces maximum portion (16.5%) of the Z-track during observation no. 9. The motion of the source along the Z-track as the function of time during this observation is shown in Figure 4. The kinematics of motion is measured in terms of the speed of source along the Z-track. We define the speed of source along the Z-track at a given value of $S_z(i)$ as $\bmod\,(S_z(i+1) - S_z(i-1))/512$ and from them the average speed is calculated for each observation (see Table 1). We find that source moves with slower average speed in NB compared to that in FB and within FB, it moves faster in the upper part.

We find broad peaked noise (PN) component with frequency $\sim 4-8$ Hz and FWHM $\sim 6-11$ Hz both in NB and FB (Figure 5). This feature is present between $S_z$=1.83 to $S_z$=2.79, which covers about 80% of the whole Z track (Table 2). However, in between there are a few points, where PN is not present and power spectrum is best fitted by simple power law representing the VLFN.

We investigate the properties of PN and VLFN as the function of position in CD. Its properties as the function of position along the Z track are shown in Figure 6. The centroid frequency and FWHM of PN does not vary systematically as the source moves from NB to FB. However, the average centroid frequency is slightly higher ($\sim 6.5$ Hz) in NB than that in FB ($\sim 5.5$ Hz). The average value of FWHM is $\sim 8$ Hz, both in NB and FB. The relative width of PN (Figure 7) fluctuate between $0.7-2.0$ in NB ($S_z$=1.8-2.0), $1-2.2$ in the lower part of FB ($S_z$=2.0-2.2) and $1.2-1.8$ in the middle part of FB ($S_z$=2.2-2.5).

The rms strength of PN ranges from $2.3\%-5.3\%$ in the NB and $1.6\%-4.4\%$ in FB. It shows increasing trend as the source moves from NB to FB and becomes highest $\sim 4.4\%$ on the 15% way up the FB. However there are fluctuation in between and this may be due to overlapping $S_z$ intervals. The rms strength decreases as the source moves further up along the FB. The average rms strength above 30% way up the FB ($S_z = 2.36-2.6$) is $\sim 2.85\%$. It is noticed that during observation no 18 and second half of the observation no. 3, mean $S_z$ values are same (2.03) but rms values of PN during these observations are significantly different (see Table 2).

The VLFN rms strength ($0.125-1$ Hz) for the first two $S_z$ values in NB is substantially higher (see Table 2). We notice that VLFN is stronger in NB compared to FB. The average value of VLFN rms in NB (excluding the first two points) is $\sim 3\%$ and in FB is $\sim 1.9\%$.

We also probe the change in properties of PN near NB/FB vertex using single continuous observations. As mentioned earlier, we have three such observations with numbers 3, 13 and 9, for which the source is present both in NB and FB (Table 1). During the observation nos. 3 & 13, source traces only a small portion of track in CD. For observation no. 3, as the source moves from NB to FB, FWHM increases from $\sim$ 5.5 Hz to $\sim$ 8 Hz. The source shows just opposite behavior during observation no. 13. Here it moves from FB to NB and the width of PN increases from $\sim$ 9 Hz to $\sim$ 11 Hz (see Table 2). During observation no. 9, the source traces significant portion of Z-track and therefore, we systematically study (i.e., using $S_z$ parameter) the properties of PN for each 256 s for this observation. Such a study shows that PN becomes broader and %rms increases as the source moves from NB to FB (Figure 8). The rms strength is found to be highest at $S_z = 2.03$ for this observation.

We find a narrow structure in power spectrum at $S_z = 2.64$ ($\nu_c \sim 3.8$ Hz and $FWHM \sim 1.7$ Hz). This feature clearly stands out in Figure 6 and 7 and we mark it by a 'square' symbol. The quality factor of this feature is $\sim$ 2.2 and therefore, this feature is a QPO, and may not be a PN. The QPO is found when the intensity of the source was decaying and then dipping after a large intensity flare (Figure 10). To find the position of the occurrence for this QPO feature, we divide the light curve at $S_z = 2.64 \pm 0.07$ in 4 intervals of 128s and create power spectra for each of these intervals. The investigation of these power spectra reveals that QPO feature is present only during the last interval, which corresponds to the intensity dip. The power spectrum during this dip is shown in Figure 9. The finding of this QPO is significant with 90% confidence level (using F-test). The %rms, centroid frequency and FWHM of the QPO during the dip are 1.72±0.3, 3.72±0.3 and 1.41±0.4 respectively.

We investigate the photon energy dependence of PN strength in FB as well as in NB. We select two $S_z$ ranges from each of these two branches, such that %rms of PN in the energy range 2-60 keV is $\geq 4\%$ for these $S_z$ ranges. Power spectra in the energy bands 2-3.5, 3.5-6.4, 6.4-9.7 and 9.7-16 keV are constructed for each of these $S_z$ ranges and while fitting the power spectra, $\nu_c$ and FWHM of PN are fixed at the values that are the best-fit values for the energy range 5.1-60 keV. We find that rms strength of PN shows a positive correlation with the photon energy, in both NB and FB (Table 3).

Finally, we compare the CDs of Cyg X-2 at low-overall and high (or, medium) intensity states with that of GX 349+2. We make use of Figure 1 of Kuulkers et al. (1999) to calculate CDs for Cyg X-2. Both GX 349+2 and Cyg X-2 (at low overall intensity) show only NB and FB. It is noticed that the transition of Cyg X-2 from high (or, medium) intensity state to low-overall intensity state causes NB/FB vertex to move towards a higher hard color value (Figure 11).

## 4. Discussion and Conclusion

In this paper, we have analyzed $\sim$ 14 days' (total good-time duration is 39.8 ks) data for the Z source GX 349+2 observed by RXTE PCA. We have computed color-color diagram (CD) and hardness-intensity diagram (HID) for this source with 256 s time average. Such a big time average has been taken in order to differentiate between the normal branch (NB) and the flaring branch (FB) in an unambiguous way, and even with it, we get enough number of points in CD and HID, as we have a huge good-time data set. We have got an extended normal branch for this source, which for almost all of the earlier observations, showed a very small normal branch (like a blob). Therefore, this is for the first time, it has been possible to compare the properties of NB and FB (and also to study the importance of NB/FB vertex) for this source with long-term good quality data.

To understand the nature of the property-variation of the source more quantitatively, we have defined rank number ($S_z$) along the Z-track (as described in section 2). We have chosen $S_z = 2$ for the NB/FB vertex and $S_z = 3$ for the end point of FB. This is to facilitate the comparison with other five Z sources (that show horizontal branch), as for these sources HB/NB vertex is generally chosen as $S_z = 1$ and NB/FB vertex as $S_z = 2$. Our analysis shows that for $\sim 79\%$ of total observation time, the source remained in the FB (see Figure 3 and Table 1). This is in accordance with the fact that GX 349+2 was almost always found to be in FB. The Z-track given in Figure 1 is actually a combination of several Z-tracks, traced by the source in different days. Figure 4 shows how the source moves in CD in the time scale of $\sim 1$ hour on a particular day. Here it moves from FB to NB (covers $\sim 17\%$ of the length of the Z-track in $\sim 30$ minutes), i.e., in the direction of decreasing 'inferred-accretion-rate'. We also see (from Table 1) that the source can move in both the directions along the Z-track.

We have found two kinds of noises in the power spectra: (1) very-low-frequency-noise (VLFN) and (2) peaked noise (PN). For all the $S_z$ values (i.e., whole of the Z-track), VLFN has been found (like another Sco-like source GX 17+2; Homan et al. 2001). We have fitted the VLFN by a power law ($0.1 - 1.0$ Hz). The index-value comes in the range $1.3 - 2.4$ for NB and $1.1 - 2.8$ for FB (these values were $0.4 - 1.5$ and $0.8 - 4.0$ for GX 17+2 respectively; Homan et al. 2001). Therefore, we see that the VLFN is much steeper in NB for GX 349+2. The corresponding percentage rms for our source varies in the range $2\% - 13.1\%$ in NB and $0.8\% - 5.9\%$ in FB (according to Homan et al. 2001, for GX 17+2 these values were $0.4\% - 0.8\%$ and $0.5\% - 1.6\%$ respectively, for the same range of frequency). Therefore, in general VLFN for GX 349+2 is much stronger. The earlier studies have shown that VLFN strength for this source increases along FB (Ponman et al. 1988; Paul et al. 2001), but for our case it does not vary systematically along this branch (see Table 2).

The broad QPO (that is to some extent similar to N/FBO of other Z sources) was found in FB of GX 349+2 for all the past observations. By the definition of QPO (quality factor $Q > 2$), this feature in power spectra may not be called a QPO. O'Neill et al. (2001) called it 'FBN', as they found it mostly in the flaring branch (they had only one $S_z$-point in NB). However we call it 'PN', as we have discovered it in both NB and FB, and its nature does not change much from one branch to another.

We, for the first time, have studied the properties of PN quantitatively (i.e., using the $S_z$ parameter) for both NB and FB. O'Neill et al. (2001) attempted it with GINGA data, but their data-quality was poor and they did not get an extended normal branch (as is mentioned earlier). As a result, they gave the ranges of centroid frequency ($\nu_c$; $4-7$ Hz), full-width-half-maximum (FWHM; $6-12$ Hz) and the maximum value of percentage rms ($\sim 3\%$) of PN for the full Z-track. Kuulkers & van der Klis (1998) did not study the variation of PN-properties with $S_z$. Therefore, they have found the values of $\nu_c$, FWHM and percentage rms in NB ($\sim 9.4$ Hz, $\sim 16$ Hz and 3% respectively) and in FB ($\sim 6$ Hz, $\sim 11$ Hz and 4% respectively), and not the ranges. In Table 2, we have displayed a detailed variation of these parameters with $S_z$. For our case, $\nu_c$ of PN comes in the range $5.0-8.3$ Hz (on average 6.5 Hz) for NB and $4.7-6.6$ Hz (on average 5.5 Hz) for FB. Therefore $\nu_c$ on average slightly decreases, as the source moves from NB to FB. This is in accordance with the earlier results (Kuulkers & van der Klis 1998). The ranges of FWHM come out to be $5.5-11.2$ Hz (for NB) and $4.8-11.0$ Hz (for FB), which is also not very different from what previous studies found. From Table 2, we see that the ranges of percentage rms are $2.3\%-5.3\%$ in NB and $1.6\%-4.4\%$ in FB. Therefore, percentage rms in general slightly decreases from NB to FB, which is contrary to the results of O'Neill et al. (2001) and Kuulkers & van der Klis (1998). However, roughly in accordance with the results of O'Neill et al. (2001), we have found that the peak value of percentage rms (in FB) occurs at $\sim 15\%$ of the way up the FB. Table 2 shows that PN appears upto 80% of the length of FB, whereas previous results (O'Neill et al. 2001) show that it occurs upto atmost 40% of FB-length. It is also possible that the variation of PN-properties depends on the direction of motion of the source along the Z-track. For example, for observation no. 3 (source moves from NB to FB) FWHM increases from 5.5 Hz to 8 Hz, and for observation no. 13 (source moves in opposite direction, i.e., from FB to NB) FWHM increases from 9 Hz to 11 Hz. Therefore, although these two observations were made on different days, it is likely that the variation of PN-properties along the Z-track is not reversible.

It is worth comparing the results of GX 349+2 with those of other two Sco-like sources, namely, GX 17+2 and Sco X-1. We have found that Sco X-1 has the highest value of $l_{\rm NB}/l_{\rm FB}$ (ratio of NB-length to FB-length) among these three sources and GX 17+2 has the lowest value (see Homan et al. 2001; Dieters & van der Klis 2000). GX 349+2 have never traced HB in the CD unlike the other two sources. But the most important difference is (as mentioned in section 1) that GX 17+2 and Sco X-1 show narrow QPOs (NBO in the normal branch and FBO in the flaring branch), whereas GX 349+2 shows a broad QPO-like structure (PN in both NB and FB). There is a possibility that these two phenomena have similar origin. This is because they have a few properties common. For example, $\nu_c$ for NBO has the value ($5-7$ Hz) similar to that for PN, and both N/FBO and PN can be fitted by Lorentzian. Besides, the strength of each of N/FBO (Sco X-1; Dieters & van der Klis 2000) and PN (GX 349+2; see Table 3) increases with increasing photon energy. In addition to that, there is no time-lag between high-energy and low-energy photons for both N/FBO and PN. However, there are several mismatch among the properties of these two. It is already mentioned that N/FBO is narrow ($Q$-value almost always is greater than 2), while PN is broad ($Q$-value is generally less than 1). In addition to that, NBO and FBO are two different QPOs, although it is widely believed that these two are same phenomena and NBO quickly transforms to FBO at the NB/FB vertex. However, there is no doubt that PN in both NB and FB is the same phenomenon. Even if NBO and FBO are the same QPO (i.e., N/FBO), there are convincing differences between it and PN. At the NB/FB vertex, NBO-frequency suddenly increases by a factor of $\sim 2$ to become FBO-frequency, and the values of $\nu_c$ and FWHM of FBO then increase along the flaring branch, while the values of these parameters for PN remain almost same throughout the Z-track. Therefore, the properties of the source that cause the track in CD to turn at the NB/FB vertex, is not probably connected to PN, whereas they must be correlated to N/FBO. Furthermore, for GX 17+2, NBO occurs upto 35% of the length of NB from the NB/FB vertex and FBO appears upto 20% of the length of FB from the same vertex (Homan et al. 2001), and these numbers are 50% and 10% respectively (Dieters & van der Klis 2000) for Sco X-2. But for GX 349+2, we see that PN is present in the whole length of NB and in the 80% length of FB. It is therefore very likely that N/FBO and PN have different origins. However, for one value (2.64) of $S_z$, we have found (see Figure 9) a narrow PN (or QPO with $Q > 2$). This may indicate some connection between N/FBO and PN, if this QPO is actually a PN (see a latter paragraph for discussion).

From the content of the previous paragraph, it may be clear that a single model for both N/FBO and PN is very difficult to formulate. According to the standard model for N/FBO (Fortner et al. 1989; Lamb 1991; Psaltis et al. 1995), radial oscillations in the optical depth of radial inflow (caused by the radiation pressure at near Eddington luminosity) produce a rocking in the X–ray spectrum, which gives rise to NBO. Although this model may be able to explain N/FBO of Sco X-1, it can not explain PN, as PN-frequency does not increase with the 'inferred accretion rate' in FB. Besides for our case, PN has been observed 80% of the way up the FB. According to the standard model, oscillations are supposed to be suppressed at such high accretion rate. An alternative model (Titarchuk et al. 2001), that identifies NBO-frequency as the spherical-shell-viscous-frequency, is also not adequate to explain PN.

A peaked noise (similar to that for GX 349+2) was observed for the Cyg-like Z source Cyg X-2 (Kuulkers et al. 1999) at low overall intensities. The noise-component extends from 2 Hz to 20 Hz in the power spectrum, peaking near $6-7$ Hz (percentage rms 3%).

To investigate whether the peaked noise components of GX 349+2 and Cyg X-2 have the same origin, we have computed the CD of GX 349+2 using the same energy ranges as given in Kuulkers et al. (1999). We have found (Figure 11) that at low overall intensities, the Z-track of Cyg X-2 looks similar to that of GX 349+2 (none of them shows HB), and although for GX 349+2, the Z-track is shifted towards higher soft color, there may be some connection between these two sources through the hard color. This is because the hard-color-value of the NB/FB vertex for Cyg X-2 increases (i.e., shifts towards (actually becomes more) that for GX 349+2), when it changes from high (or, medium) overall intensity to low overall intensity (the gradual change has not been observed). All these indicate that the nature of the X–ray emitting components of GX 349+2 and that of Cyg X-2 (at low overall intensities) may be similar to some extent and the intensity in the energy range $6.4 - 16.0$ keV (used to calculate hard color) may be an important clue to understand such similarity. If the origin of peaked noise components of the two sources are actually same, then it may be possible that such noise components are originated in the region, which produces most of the luminosity in $6.4 - 16.0$ keV energy range. This is supported by the fact that the rms strength of PN is higher in the energy range $7 - 16$ keV compared to that in a lower energy range (see Table 3). However, as the CD represents a rough spectral behavior, to establish a connection between the two PN components, it is essential to compare the energy spectra of these two sources in details.

We have already mentioned that for $S_z = 2.64$, we have found a QPO for GX 349+2. This is, to our knowledge, the first low-frequency-QPO seen for this source (if we do not consider PN as QPO). The FBOs with $\nu_c \geq 10$ Hz have been observed from other five Z sources. But the QPO we see for this source has a much lower frequency ($\nu_c = 3.8$ Hz) and therefore, it is certainly not a FBO. There is a possibility that it has the same origin as of PN (because at the time when it occurs, PN does not occur), but appears with higher quality factor. However from Figure 6, we see that its properties (marked by a 'square') are very different from the properties of PN. Therefore it is unlikely that these two have the same origin. A similar 26 Hz QPO has been observed from Cyg X-2 during the intensity dip (Kuulkers & van der Klis 1995). There, the authors suggest that a thick torus like structure, formed due to puffing up of inner disk at large accretion rate may obscure the inner emission region from our line of sight and oscillation in such a obscuring torus may provide the explanation for this type of the QPOs. This model requires a higher inclination of the source. The inclination of Cyg X-2 is $\sim 60°$ (Orosz and Kuulkers 1999). If the QPO seen for GX 349+2 is also due to the oscillation in such a thick torus, then the inclination of this source should be close to that of Cyg X-2.

As mentioned in section 1, GX 349+2 is not a well-observed source and certainly is the most poorly understood one among the Z sources. However, it shows interesting phenomenon like PN (which is probably also observed from Cyg X-2 for a particular position of Z-track in CD). Therefore, the study of this source may be very important to understand the physics behind the tracks traced by the Z sources in CD and HID. The detailed study of GX 349+2 in this paper will be useful for this purpose, as well as will help to formulate a correct theoretical model for PN.

*Acknowledgements.* We deeply acknowledge P. Sreekumar for providing all the facilities and for detailed discussions. We thank A.R. Rao for reading the manuscript and giving valuable suggestions. We also thank Dipankar Bhattacharya for his help and the Director of Raman Research Institute for the facilities provided.

## References

Boyle, C.B., Fabian, A.C., & Guilbert, P.W. 1986, Nature, 319, 648

Dieters, S.W., & van der Klis, M. 2000, MNRAS, 311, 201

Fortner, B., Lamb, F.K., & Miller, G.S. 1989, Nature, 342, 775

Hameury, J.M., King, A.R., & Lasota, J.P. 1985, Nature, 317, 597

Hasinger, G. & van der Klis, M. 1989, A&A, 225, 79

Hasinger, G., van der Klis, M., Ebisawa, K., Dotani, T., & Mitsuda, K. 1990, A&A, 235, 131

Hertz, P., Vaughan, B., Wood, K.S., et al. 1992, ApJ, 396,201

Homan, J., van der Klis, M., Jonker, P.G., Wijnands, R., Kuulkers, E., Mendez, M., & Lewin, W.H.G. 2001, astro-ph/0104323

Kuulkers, E., Homan, J., van der Klis, M., Lewin, W.H.G., & Mendez, M. 2001, astro-ph/0105386

Kuulkers, E. & van der Klis, M. 1995, A&A, 303, 801

Kuulkers, E. & van der Klis, M. 1996, A&A, 314, 567

Kuulkers, E., & van der Klis, M. 1998, A&A, 332, 845

Kuulkers, E., van der Klis, M., Vaughan, B.A. 1996, A&A, 311, 197

Kuulkers, E., van der Klis, M., Oosterbroek, T., Asai, K., Dotani, T., van Paradijs, J.,Lewin, W.H.G. 1994, A&A, 289, 795

Kuulkers, E., van der Klis, M, Oosterbroek, T., van Paradijs, J., Lewin, W.H.G. 1997, MNRAS, 287, 495

Kuulkers, E., Wijnands, R., & van der Klis, M. 1999, MNRAS, 308, 485

Lamb, F.K. 1991, Unified Model of X–ray Spectra and QPOs in Low Mass Neutron Star Binaries. In: Ventura J., Pines D. (eds.), Neutron Stars: Theory and Observations, Dordrecht, Kluwer Academic Publishers, p. 445

Lamb, F.K., Shibazaki, N., Alpar, M.A., & Shaham, J. 1985, Nature, 317, 681

O'Neill, P. M., Kuulkers, E., Sood, R. K., & Dotani, T. 2001, A&A, 370, 479

Orosz, J.A., Kuulkers, E. 1999, MNRAS, 305, 132

Ponman, T.J., Cooke, B.A., & Stella, L. 1988, MNRAS, 231, 999

Psaltis, D., Lamb, F.K., & Miller, G.S. 1995, ApJ, 454, L137

Smale, A.P. 1998, ApJ, 498, L141.

Titarchuk, L.G., Bradshaw, C.F., Geldzahler, B.J., & Fomalont, E.B. 2001, astro-ph/0105559

van der Klis, M. 1995, Rapid Aperiodic variability in X-ray binaries. In: Lewin, W.H.G., van Paradijs, J., van den Heuvel, E.P.J. (eds), X-ray binaries, Cambridge University Press, Cambridge, p. 252

van der Klis, M. 2000, ARA&A, 38, 717

Zhang, W., Strohmayer, T.E., Swank, J.H. 1998, ApJ, 500, 167

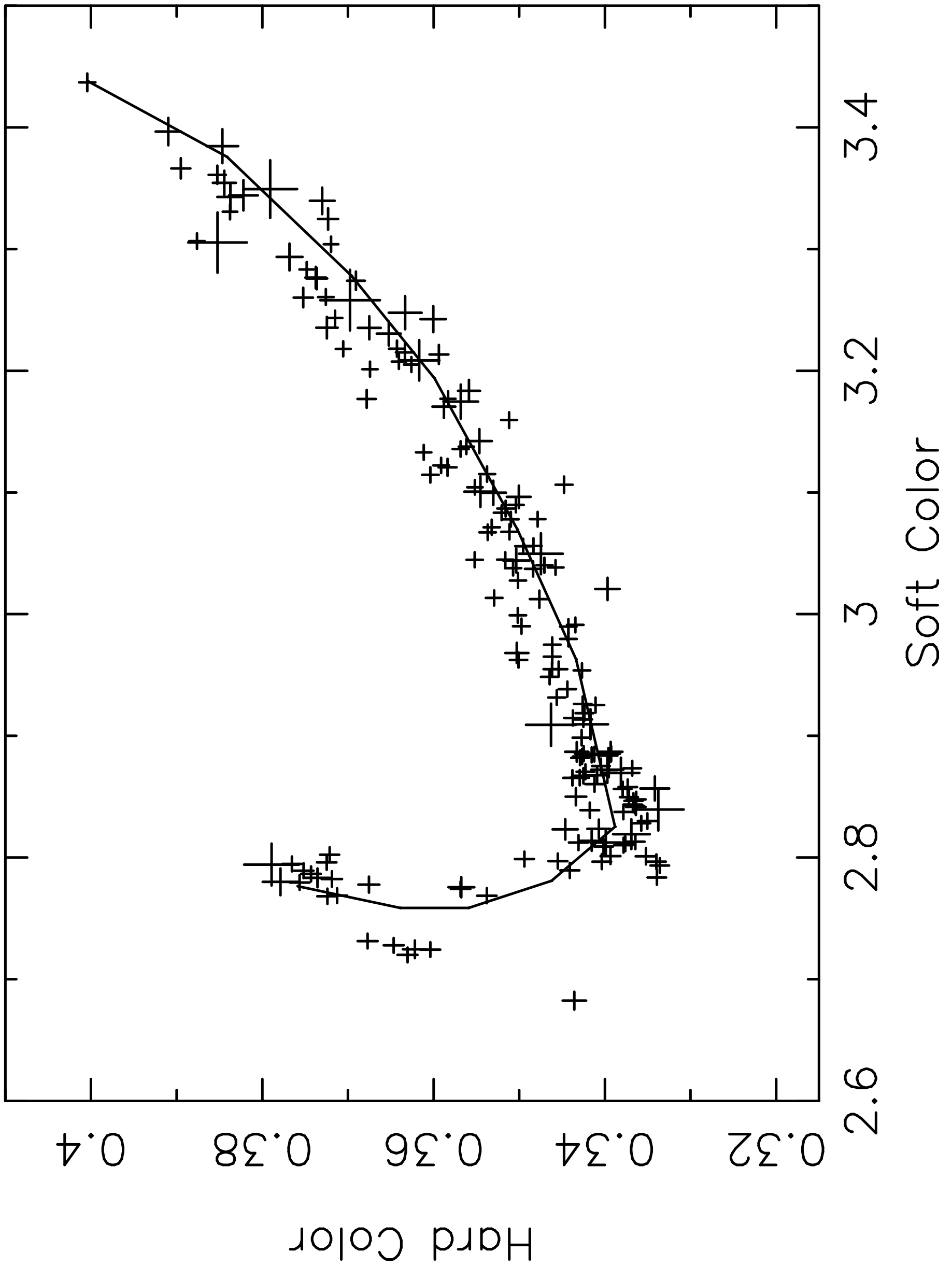

**Fig. 1.** Color-Color diagram for observations from September 29 to October 13, 1998. Soft color is the ratio of count rates in energy band 3.5-6.4 keV and 2-3.5 keV and hard color is the count-rate-ratio in energy bands 9.7-16 keV and 6.4-9.7 keV. Each point corresponds to a 256 s bin size.

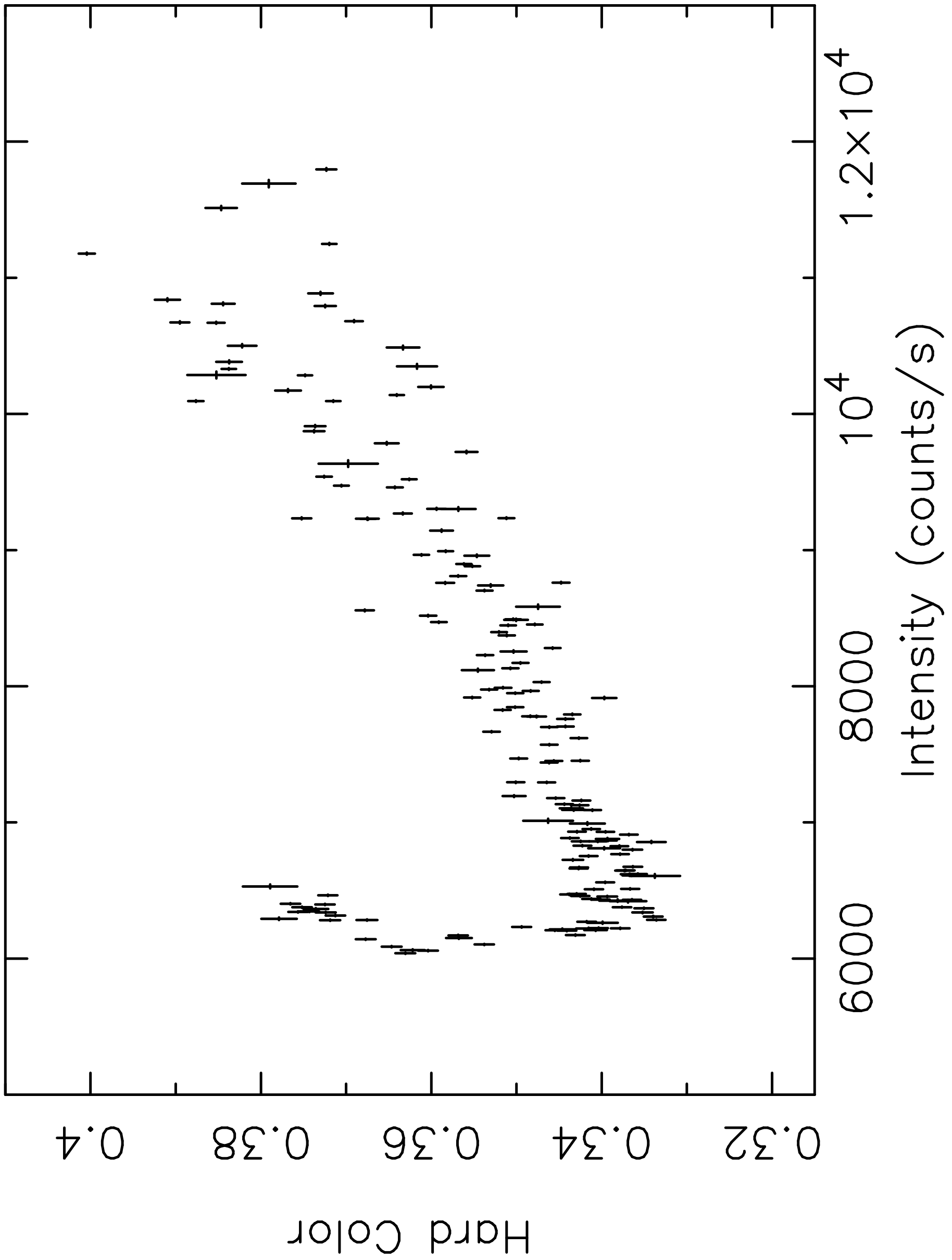

**Fig. 2.** Hardness-Intensity diagram for observations from September, 29 to October 13, 1998. The intensity is the count rate in energy band 2-16 keV.

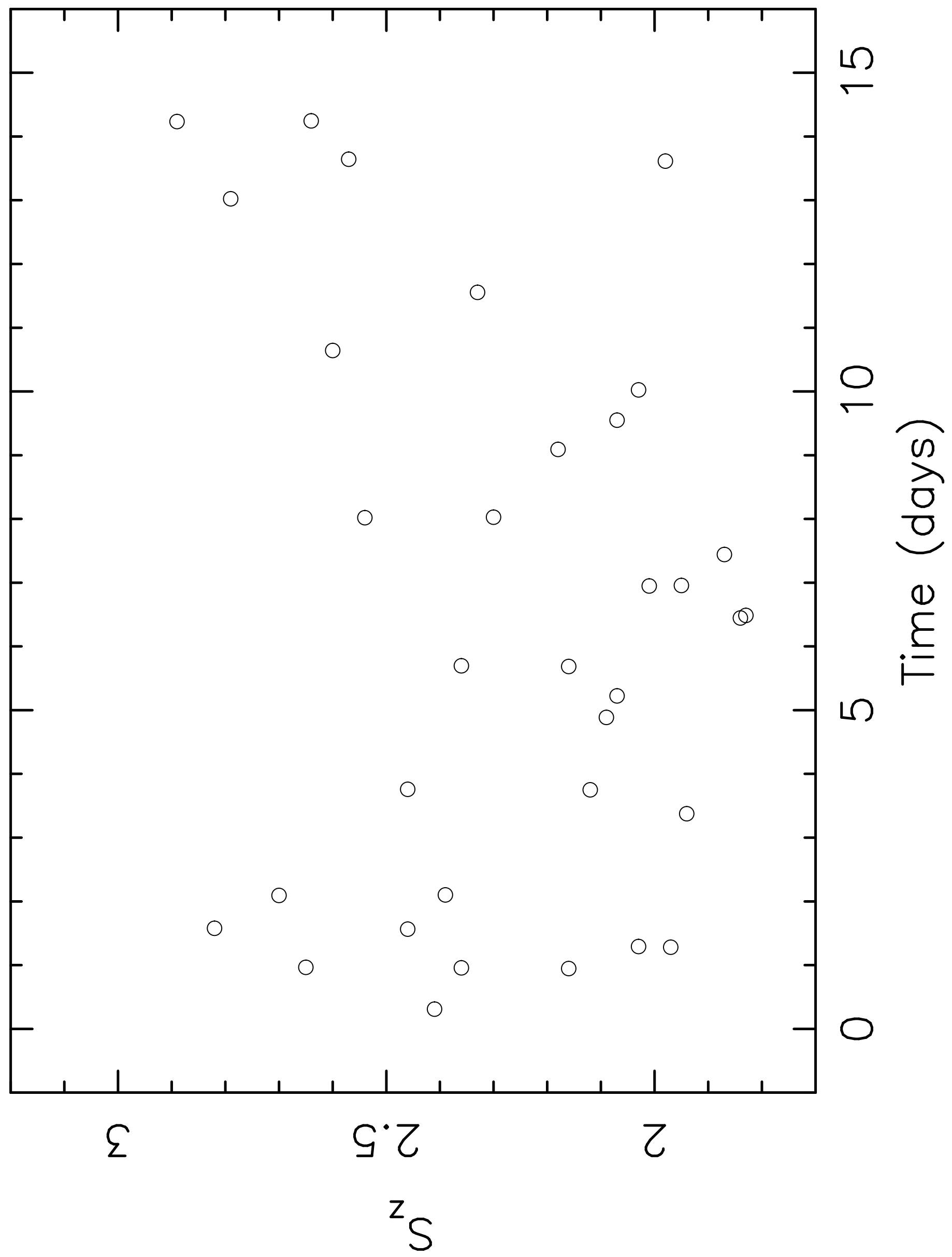

**Fig. 3.** Variation of $S_z$ parameter on time scale of several days. Start date and time for this plot are 29/09/98 and 07:07:26 UT respectively .

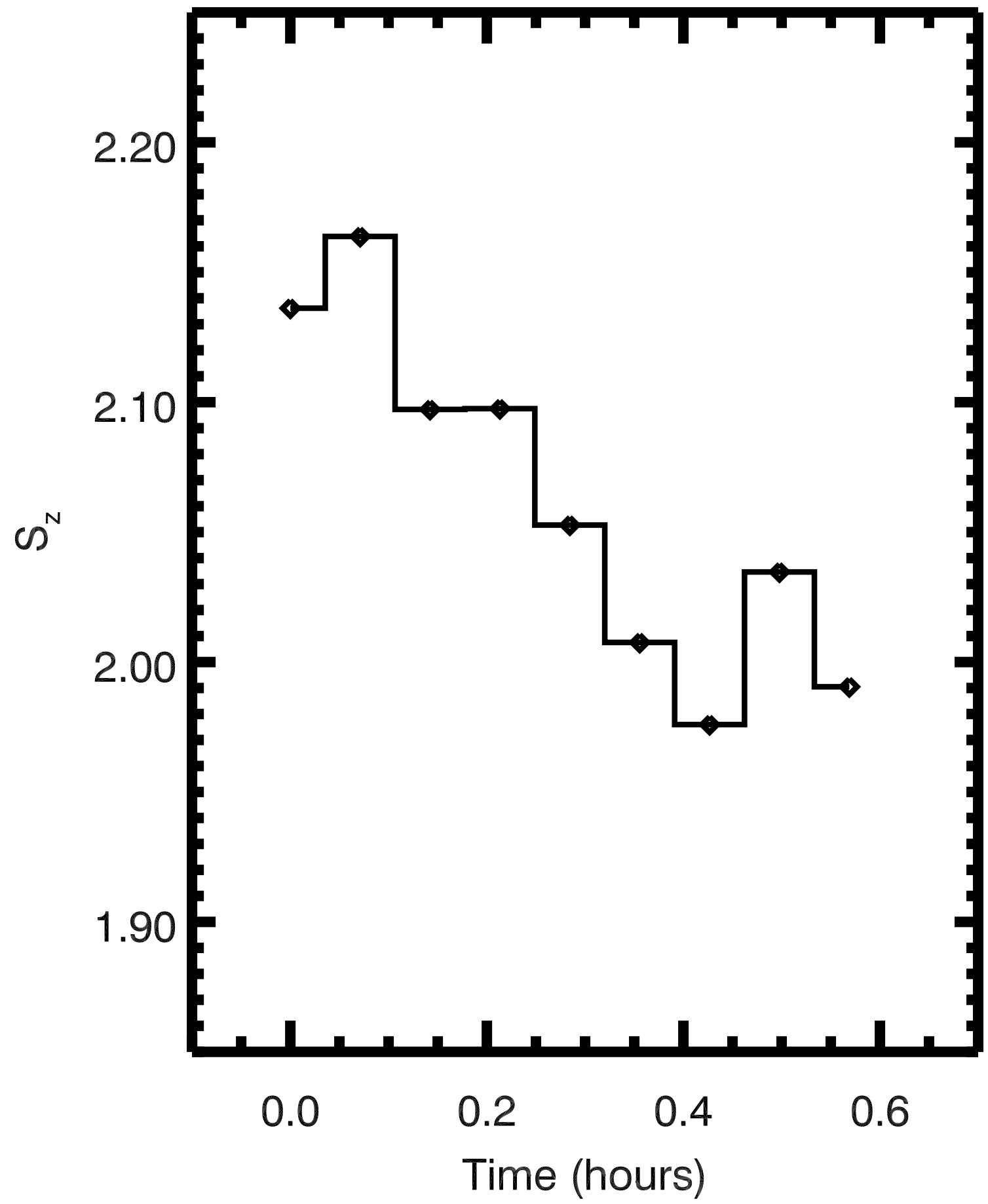

**Fig. 4.** Variation of $S_z$ parameter for observation no. 9. Start time for this plot is 5:00:48 UT. Each point has a bin size 256 s.

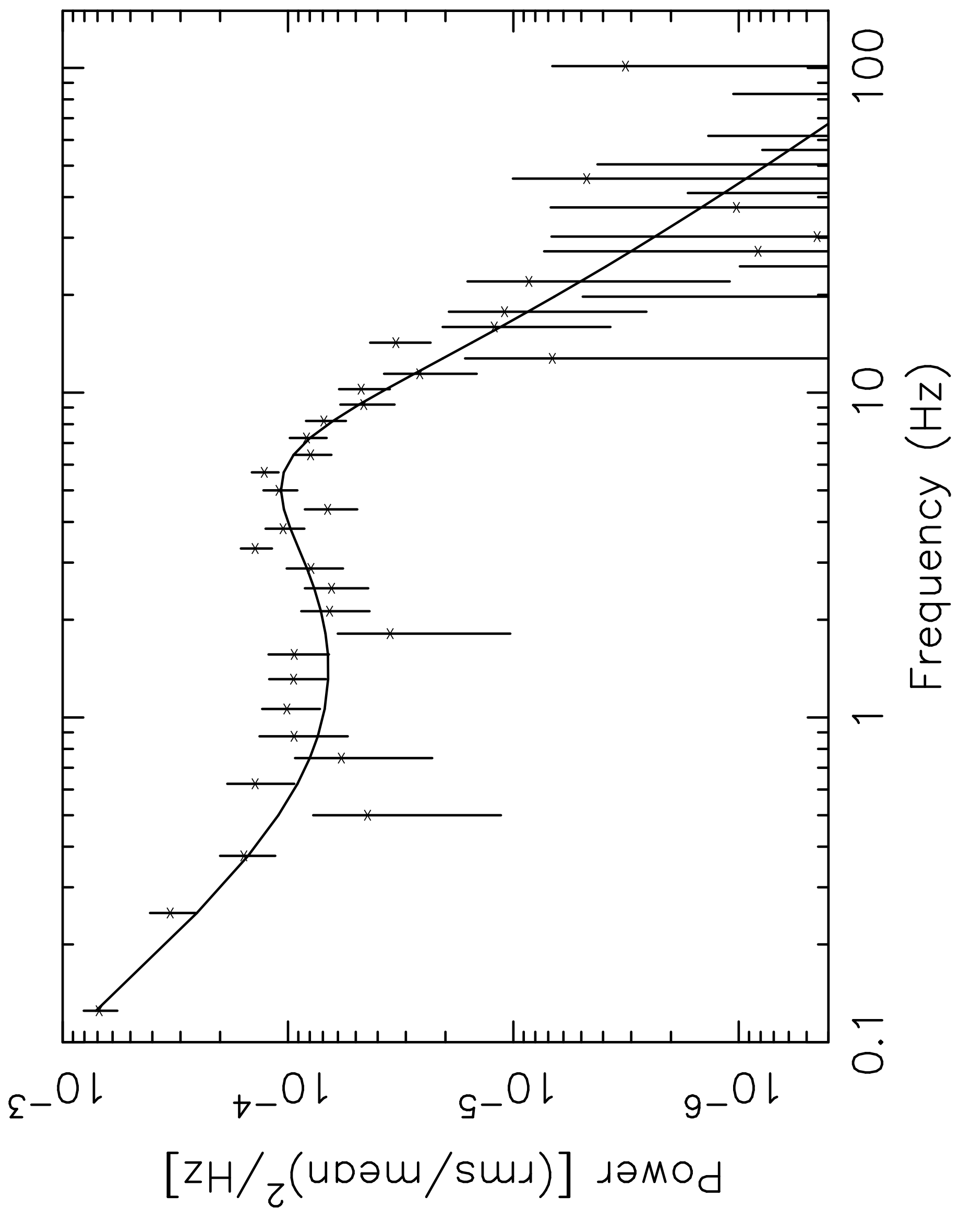

**Fig. 5.** Power spectrum in the energy band 5.1-60 keV at $S_z = 2.36 \pm 0.09$. The solid line shows the best fitted curve (power-law+Lorentzian).

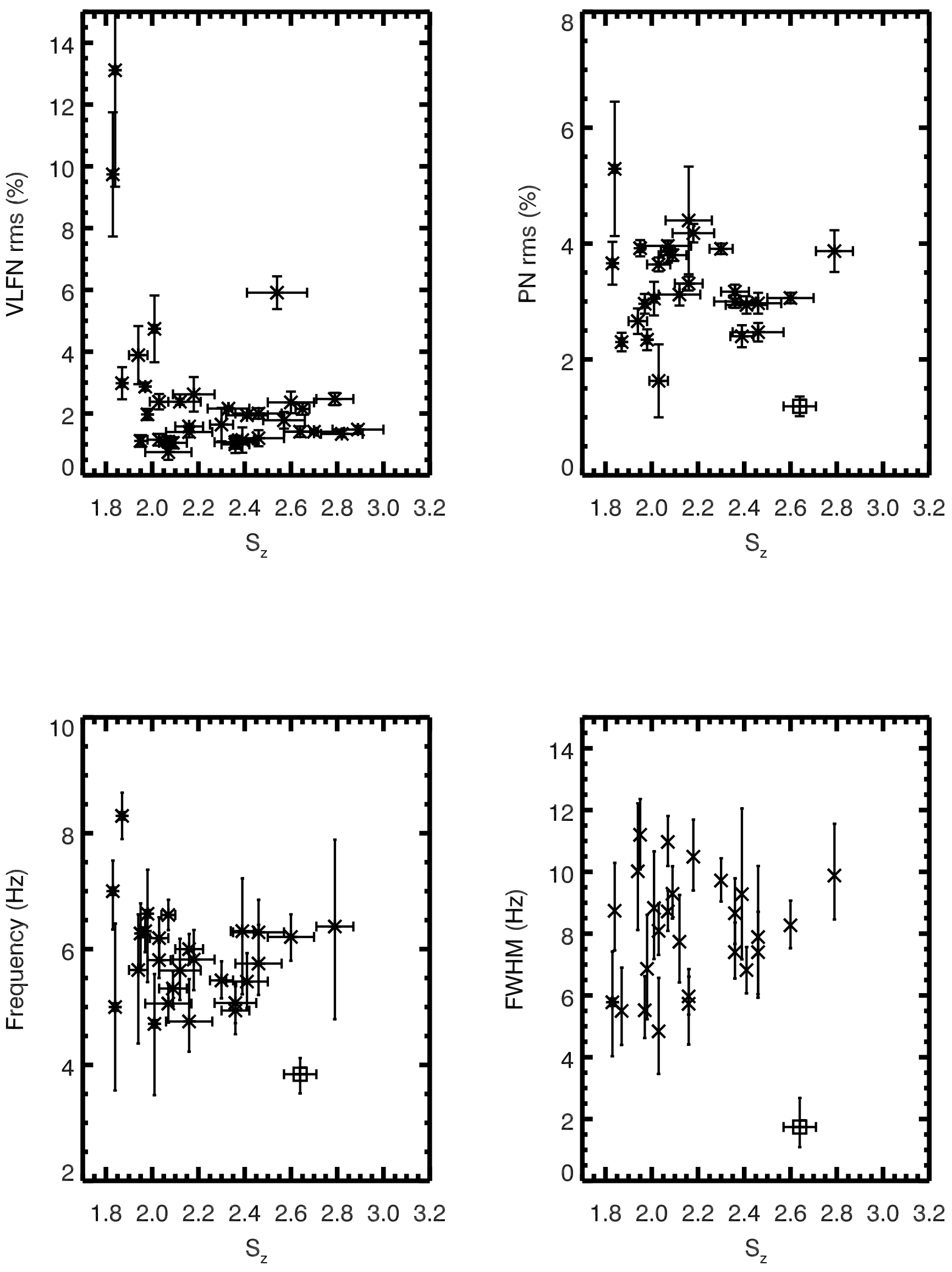

**Fig. 6.** Properties of VLFN (%rms) and PN (%rms, centroid frequency and FWHM) as function of $S_z$. The rms strength of VLFN is calculated by integrating the power-law-curve in the frequency range 0.125-1 Hz. The symbol 'square' is for $S_z = 2.64$, for which a QPO has been observed.

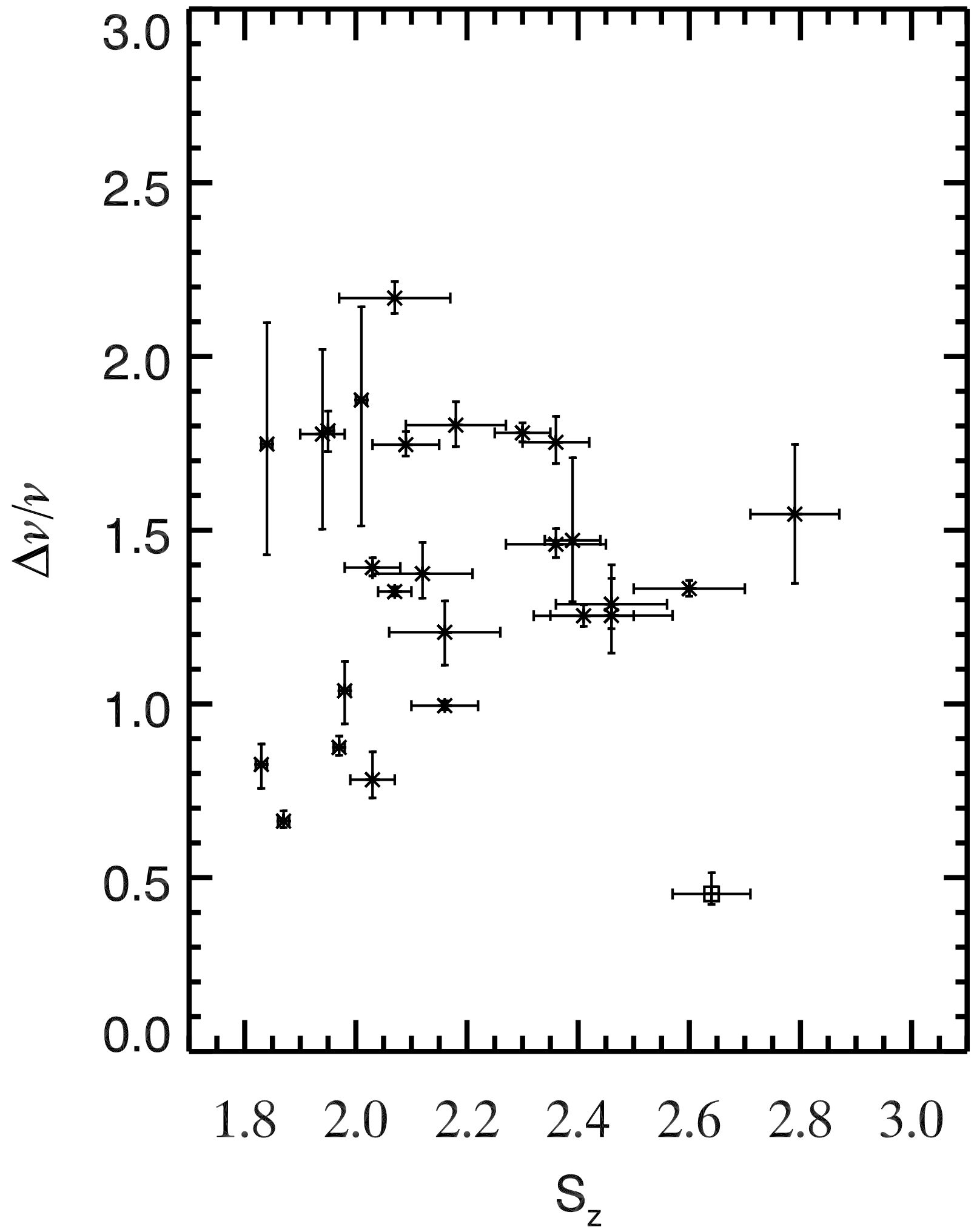

**Fig. 7.** Variation of relative width of PN with the position along the Z-track. The symbol 'square' bears the same meaning as in figure 6.

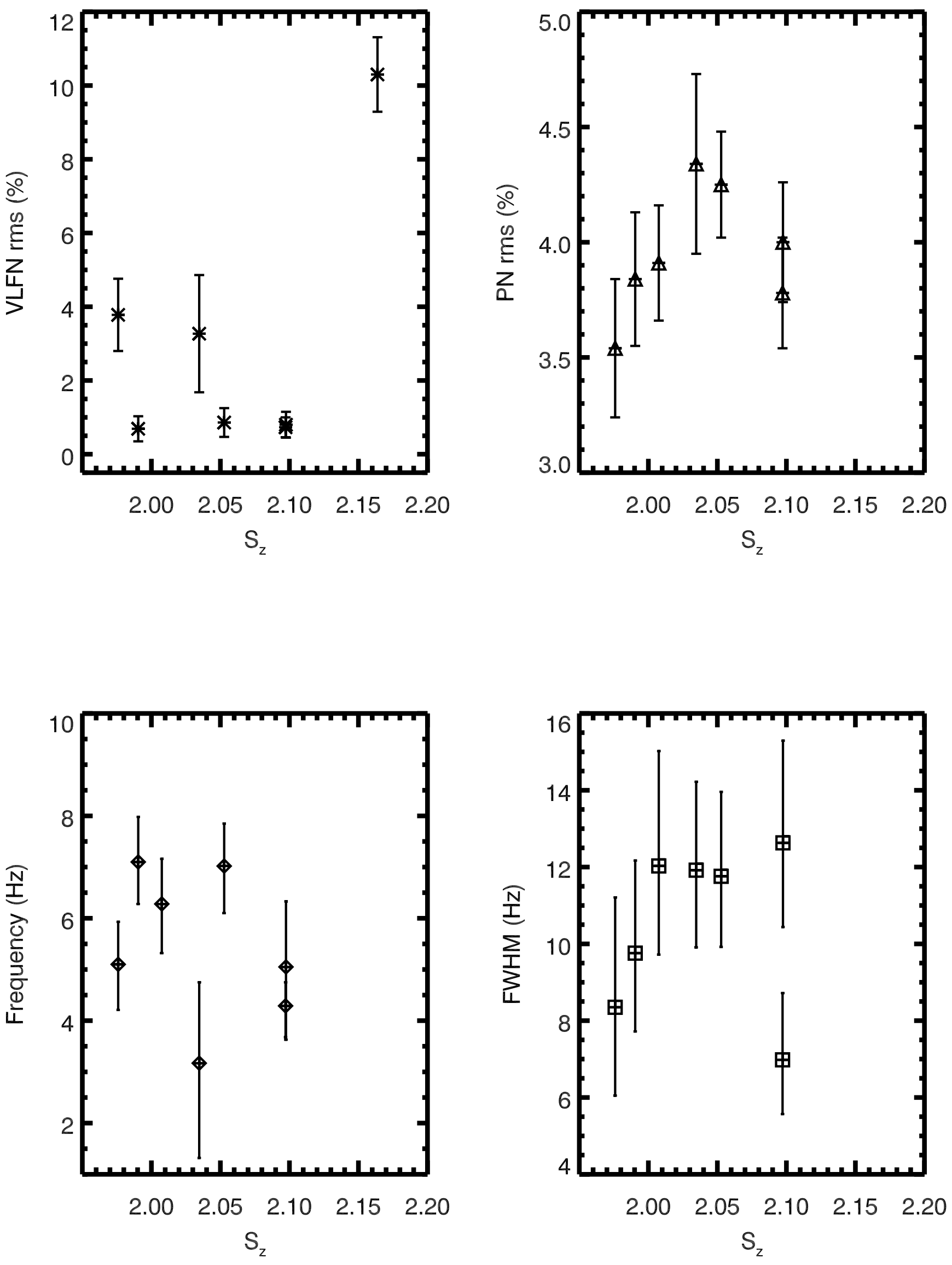

**Fig. 8.** Properties of PN and VLFN as functions of $S_z$ for Observation no. 9.

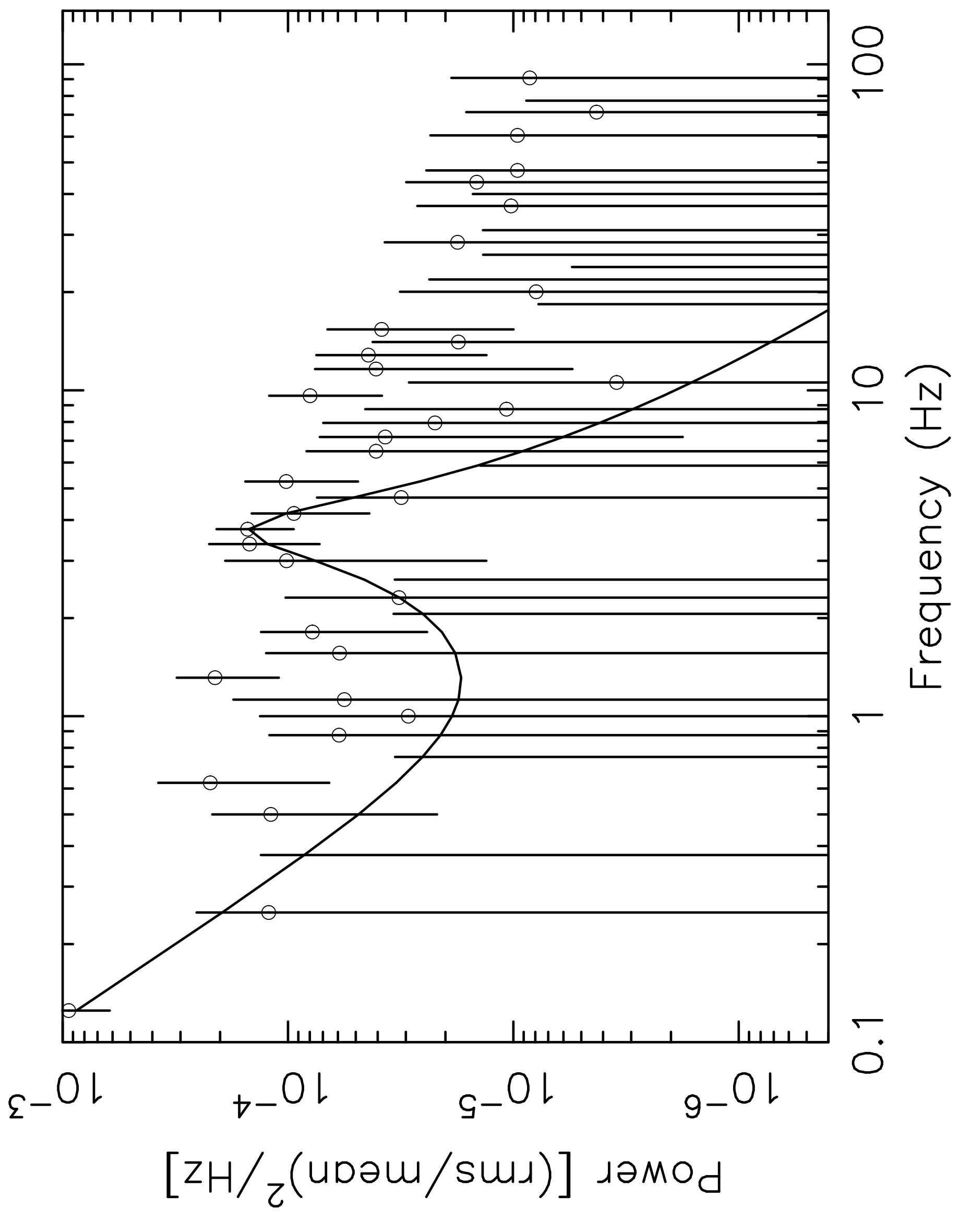

**Fig. 9.** Power spectrum in energy band 5.1-60 keV during the intensity dip ($S_z = 2.64$). The solid line shows best-fitted-curve (power-law+Lorentzian). The narrow feature in the power spectrum is the QPO at $\sim 3.8$ Hz.

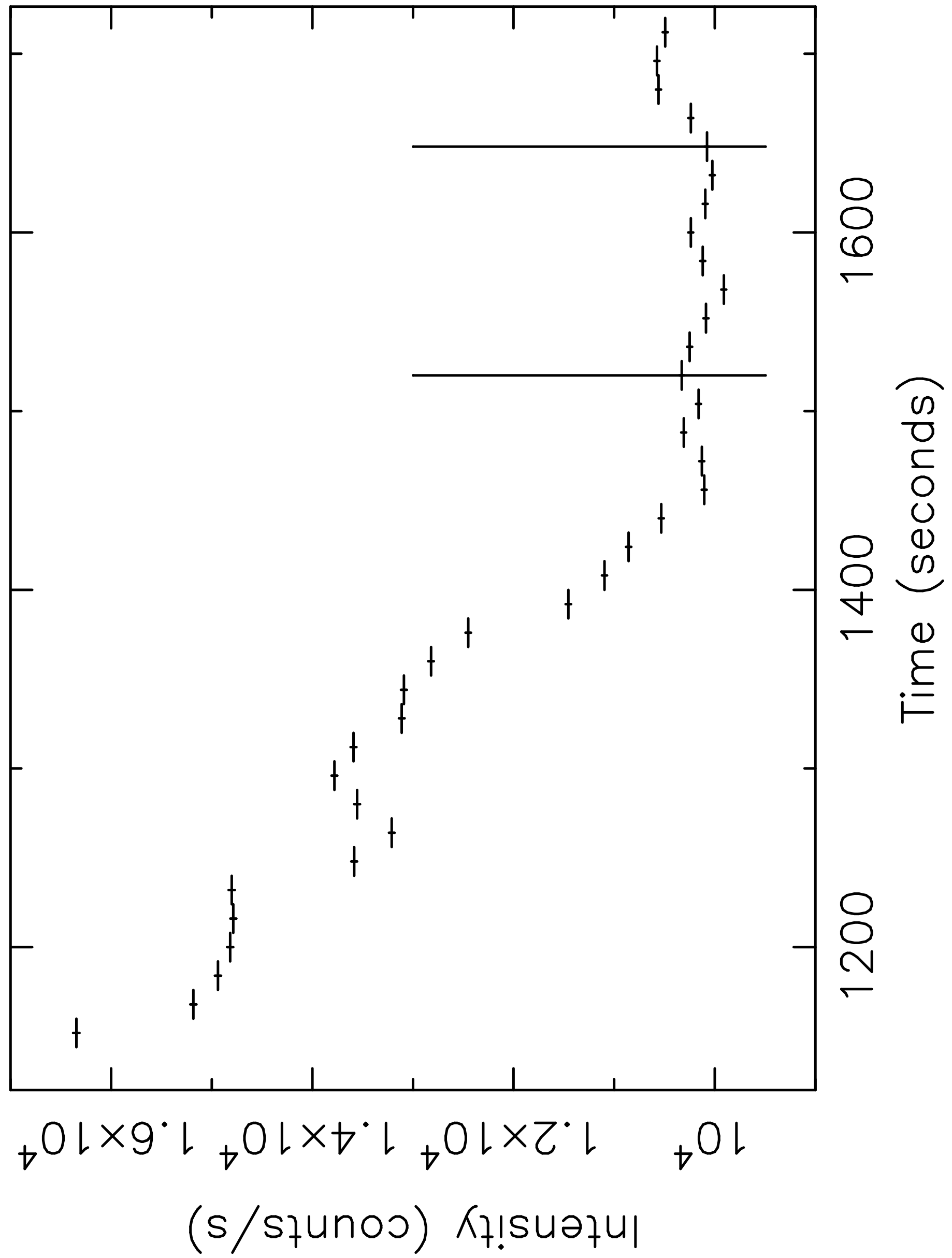

**Fig. 10.** The light curve of GX 349+2 at $S_z = 2.64 \pm 0.07$ (observation no. 24) with a 16 s time resolution. A clear intensity dip is observed during this period. A low-frequency QPO is found in the region between the two vertical solid lines. The start time of this light curve corresponds to 5:29:59 UT.

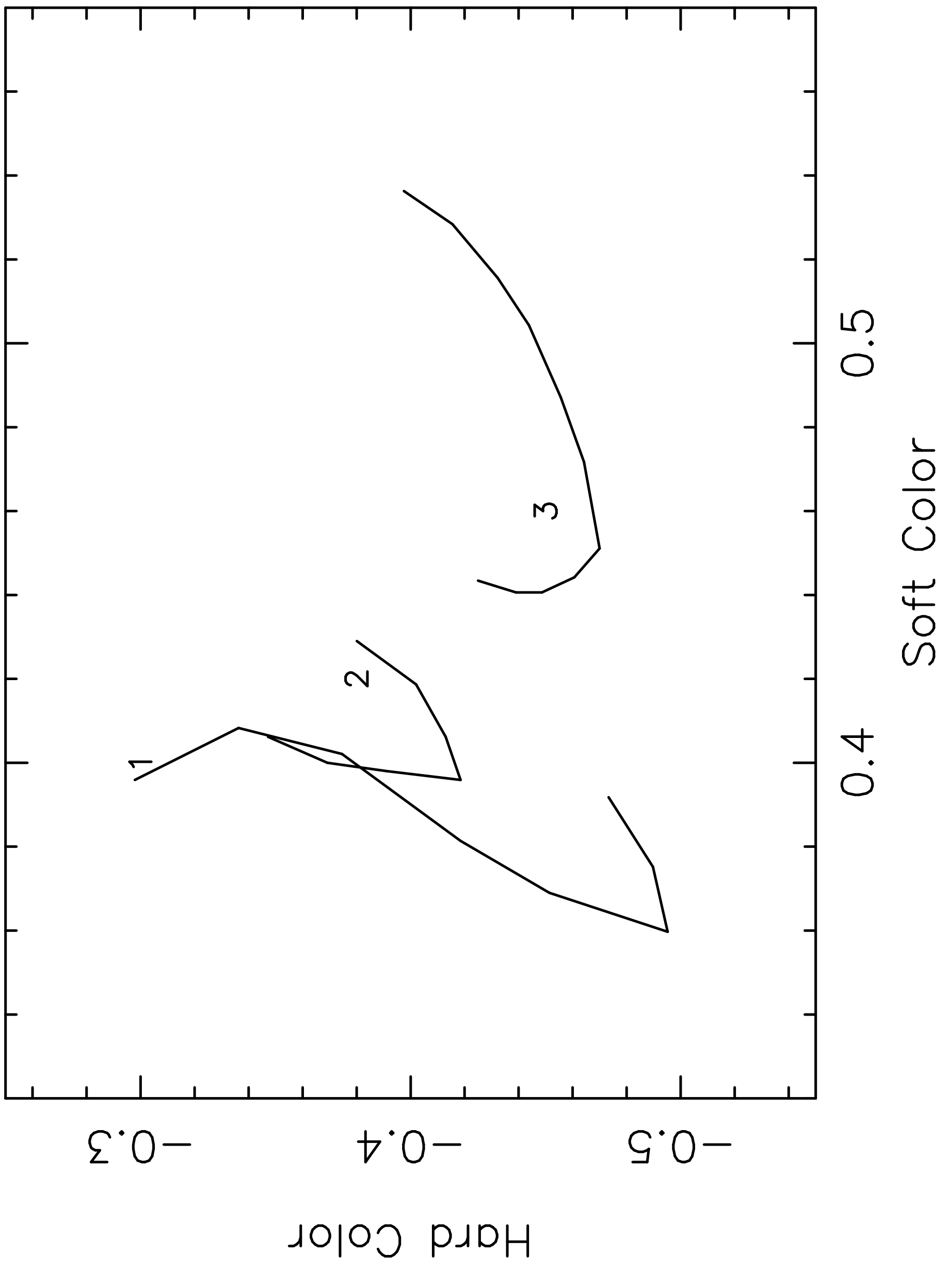

**Fig. 11.** Comparison between the spectral behaviors of Cyg X-2 in low intensity state and GX 349+2. The curve with label '1' is Z-track for Cyg X-2 in high/medium intensity state and that with label '2' is Z-track in low intensity state. These tracks have been generated using figure 1 of Kuulkers et al. (1999). The curve with label '3' is the Z-track for GX 349+2. Here the definitions of soft color and hard color are slightly different, i.e., they are the logarithms of soft color and hard color (respectively) used in figure 1.

| Observation number | Observation ID | Date of observation | Start time UT | End time UT | Good time duration (s) | State | $< V_z >$ ($\times 10^{-4}$) | % of Z track |
|---|---|---|---|---|---|---|---|---|
| 1 | 30043-01-01-00 | 29/09/98 | 07:07:28 | 07:40:07 | 1938 | mFB | 1.166 | 16.2 |
| 2 | 30043-01-02-00 | 29/09/98 | 22:40:16 | 23:13:07 | 1969 | lFB→uFB | 3.611 | 54 |
| 3 | 30043-01-03-00 | 30/09/98 | 06:39:00 | 07:11:07 | 1924 | NB→lFB | 0.586 | 10 |
| 4 | 30043-01-04-00 | 30/09/98 | 13:23:28 | 13:56:07 | 1700 | mFB→uFB | 2.591 | 35 |
| 5 | 30043-01-05-00 | 01/10/98 | 01:59:28 | 02:32:07 | 1762 | uFB→mFB | 3.125 | 42 |
| 6 | 30043-01-07-00 | 02/10/98 | 08:45:20 | 09:18:07 | 1967 | NB | 0.400 | 6.1 |
| 7 | 30043-01-08-00 | 02/10/98 | 17:52:44 | 18:22:07 | 1750 | lFB→uFB | 4.791 | 67.8 |
| 8 | 30043-01-10-00 | 03/10/98 | 21:04:16 | 21:30:07 | 1392 | lFB | 0.716 | 11.3 |
| 9 | 30043-01-11-00 | 04/10/98 | 05:04:23 | 05:37:07 | 1962 | lFB→NB | 0.997 | 16.5 |
| 10 | 30043-01-12-00 | 04/10/98 | 16:17:02 | 16:49:07 | 1860 | lFB→mFB | 2.063 | 22.6 |
| 11 | 30043-01-13-00 | 05/10/98 | 10:37:06 | 10:53:07 | 881 | NB | 0.077 | 0.5 |
| 12 | 30043-01-13-00 | 05/10/98 | 11:28:32 | 11:54:07 | 1515 | NB | 0.077 | 0.5 |
| 13 | 30043-01-14-00 | 05/10/98 | 22:41:02 | 23:07:07 | 1477 | lFB→NB | 0.672 | 5.2 |
| 14 | 30043-01-15-00 | 06/10/98 | 10:21:57 | 10:53:07 | 1678 | NB | 0.202 | 2.6 |
| 15 | 30043-01-16-00 | 07/10/98 | 00:17:02 | 00:50:07 | 1965 | uFB→mFB | 2.944 | 27.8 |
| 16 | 30043-01-18-00 | 08/10/98 | 01:52:35 | 02:33:06 | 2287 | mFB | 1.538 | 14.7 |
| 17 | 30043-01-19-00 | 08/10/98 | 13:04:35 | 13:36:07 | 879 | lFB | 0.733 | 2.6 |
| 18 | 30043-01-20-00 | 09/10/98 | 00:16:10 | 00:51:07 | 2056 | lFB | 0.667 | 6.9 |
| 19 | 30043-01-21-00 | 09/10/98 | 15:06:34 | 15:39:07 | 1792 | uFB | 2.938 | 21.7 |
| 20 | 30043-01-23-00 | 10/10/98 | 13:03:49 | 13:36:07 | 1933 | mFB | 1.003 | 10.43 |
| 21 | 30043-01-26-00 | 12/10/98 | 00:15:39 | 00:51:07 | 1959 | uFB | 3.111 | 11.30 |
| 22 | 30043-01-27-00 | 12/10/98 | 14:38:24 | 14:49:07 | 626 | NB | 0.250 | 1.7 |
| 23 | 30043-01-27-00 | 12/10/98 | 15:16:32 | 15:37:07 | 1154 | mFB→uFB | 2.112 | 16.52 |
| 24 | 30043-01-28-00 | 13/10/98 | 05:29:42 | 05:59:07 | 1676 | uFB | 3.388 | 25.2 |

**Table 1.** Details of the observation: Columns 4 and 5 are respectively the start and the stop time for science-event-mode-data. Column 6 gives the amount of good time available in this time interval. Column 7 shows the position of the source on the Z-track and in column 8 the average speed (defined in section 3) of the source along the track is displayed. The last column is the percentage of the full Z-track covered by the source

| Obs. no. | $S_z$ | VLFN | | PN | | |
|---|---|---|---|---|---|---|
| | | Index | rms (%) | Frequency (Hz) | FWHM (Hz) | rms (%) |
| 12 | 1.83± 0.01 | $2.10^{+0.36}_{-0.46}$ | 9.74±2.01 | $7.00^{+0.53}_{-0.66}$ | $5.78^{+1.64}_{-1.75}$ | 3.66±0.37 |
| 11 | 1.84± 0.01 | $2.41^{+0.46}_{-0.55}$ | 13.11±3.77 | $5.00^{+1.44}_{-1.44}$ | $8.74^{+1.55}_{-1.2}$ | 5.29±1.16 |
| 14 | 1.87±0.01 | $1.99^{+0.39}_{-0.59}$ | 2.98±0.52 | $8.30^{+0.40}_{-0.40}$ | $5.50^{+1.40}_{-1.10}$ | 2.30±0.16 |
| 6 | 1.94±0.04 | $1.86^{+0.60}_{-0.60}$ | 3.89±0.94 | $5.64^{+0.96}_{-1.27}$ | $10.02^{+2.20}_{-1.90}$ | 2.66±0.22 |
| $13^b$ | 1.95±0.01 | $1.41^{+0.16}_{-0.25}$ | 1.11±0.19 | $6.27^{+0.52}_{-0.59}$ | $11.20^{+1.16}_{-1.22}$ | 3.92±0.14 |
| $3^a$ | 1.97±0.01 | $1.25^{+0.06}_{-0.06}$ | 2.87±0.13 | $6.31^{+0.36}_{-0.36}$ | $5.52^{+1.1}_{-0.90}$ | 2.96±0.17 |
| 22 | 1.98±0.01 | $1.44^{+0.10}_{-0.11}$ | 1.97±0.18 | $6.61^{+0.76}_{-1.18}$ | $6.86^{+1.75}_{-1.63}$ | 2.34±0.18 |
| $13^a$ | 2.01±0.01 | $1.88^{+0.80}_{-0.80}$ | 4.74±1.08 | $4.71^{+0.86}_{-1.23}$ | $8.83^{+1.83}_{-1.65}$ | 3.05±0.29 |
| $3^b$ | 2.03±0.05 | $1.51^{+0.15}_{-0.20}$ | 1.15±0.19 | $5.81^{+0.30}_{-0.31}$ | $8.09^{+0.88}_{-0.78}$ | 3.64±0.12 |
| 18 | 2.03±0.04 | $1.81^{+0.16}_{-0.16}$ | 2.38±0.25 | $6.19^{+0.36}_{-0.40}$ | $4.84^{+1.73}_{-1.38}$ | 1.63±0.63 |
| 17 | 2.07±0.03 | $1.53^{+0.14}_{-0.19}$ | 1.09±0.17 | $6.59^{+0.26}_{-0.26}$ | $8.72^{+0.68}_{-0.63}$ | 3.86±0.19 |
| 9 | 2.07±0.10 | $1.90^{+0.35}_{-0.59}$ | 0.75±0.25 | $5.06^{+0.33}_{-0.33}$ | $10.97^{+0.84}_{-0.78}$ | 3.96±0.09 |
| 8 | 2.09±0.06 | $1.50^{+0.15}_{-0.21}$ | 1.05±0.18 | $5.32^{+0.30}_{-0.31}$ | $9.29^{+0.89}_{-0.80}$ | 3.80±0.10 |
| $7^a$ | 2.12±0.09 | $1.14^{+0.05}_{-0.06}$ | 2.38±0.10 | $5.63^{+0.55}_{-0.51}$ | $7.74^{+1.51}_{-1.32}$ | 3.12±0.19 |
| $2^a$ | 2.16±0.10 | $1.28^{+0.18}_{-0.18}$ | 1.40±0.17 | $4.75^{+0.73}_{-0.52}$ | $5.73^{+1.12}_{-1.32}$ | 4.40±0.93 |
| $10^a$ | 2.16±0.06 | $1.12^{+0.04}_{-0.04}$ | 1.58±0.08 | $6.00^{+0.26}_{-0.26}$ | $5.97^{+0.64}_{-0.59}$ | 3.31±0.12 |
| 16 | 2.18±0.09 | $1.85^{+0.28}_{-0.32}$ | 2.62±0.56 | $5.82^{+0.51}_{-0.53}$ | $10.49^{+1.20}_{-1.09}$ | 4.18±0.16 |
| $15^b$ | 2.30±0.05 | $2.00^{+0.38}_{-0.59}$ | 1.64±0.54 | $5.46^{+0.33}_{-0.31}$ | $9.72^{+0.72}_{-0.68}$ | 3.91±0.09 |
| 20 | 2.33±0.09 | $1.78^{+0.13}_{-0.13}$ | 2.16±0.07 | - | - | - |
| $2^b$ | 2.36±0.09 | $1.66^{+0.20}_{-0.26}$ | 1.07±0.21 | $5.07^{+0.33}_{-0.35}$ | $7.40^{+.96}_{-0.85}$ | 3.00±0.11 |
| $10^b$ | 2.36±0.06 | $1.85^{+0.29}_{-0.41}$ | 1.00±0.28 | $4.94^{+0.42}_{-0.41}$ | $8.66^{+1.1}_{-0.99}$ | 3.17±0.12 |
| $5^b$ | 2.39±0.05 | $2.84^{+0.51}_{-0.63}$ | 1.14±0.41 | $6.31^{+0.91}_{-1.09}$ | $9.28^{+2.77}_{-2.11}$ | 2.40±0.19 |
| 1 | 2.41±0.09 | $1.37^{+0.06}_{-0.07}$ | 1.94±0.08 | $5.44^{+0.49}_{-0.46}$ | $6.82^{+0.75}_{-0.75}$ | 2.94±0.15 |
| $7^b$ | 2.46±0.11 | $1.99^{+0.24}_{-0.30}$ | 1.20±0.26 | $6.29^{+0.56}_{-0.53}$ | $7.89^{+2.3}_{-1.96}$ | 2.47±0.16 |
| $4^a$ | 2.46±0.10 | $1.41^{+0.08}_{-0.10}$ | 2.00±0.17 | $5.75^{+0.67}_{-0.54}$ | $7.40^{+1.31}_{-1.36}$ | 2.97±0.18 |
| $15^a$ | 2.54±0.13 | $1.83^{+0.10}_{-0.10}$ | 5.91±0.53 | - | - | - |
| 23 | 2.57±0.09 | $1.81^{+0.21}_{-0.32}$ | 1.78±0.27 | - | - | - |
| 19 | 2.60±0.10 | $1.92^{+0.25}_{-0.27}$ | 2.36±0.35 | $6.21^{+0.39}_{-0.41}$ | $8.27^{+0.80}_{-0.74}$ | 3.06±0.09 |
| $24^b$ | 2.64±0.07 | $1.55^{+0.17}_{-0.14}$ | 1.41±0.17 | $3.84^{+0.28}_{-0.33}$ | $1.74^{+0.94}_{-0.65}$ | 1.19±0.17 |
| $2^c$ | 2.65±0.03 | $1.39^{+0.10}_{-0.13}$ | 2.15±0.19 | - | - | - |
| $5^a$ | 2.70±0.13 | $1.35^{+0.06}_{-0.07}$ | 1.41±0.07 | - | - | - |
| 21 | 2.79 ±0.08 | $1.23^{+0.09}_{-0.09}$ | 2.47±0.20 | $6.39^{+1.5}_{-1.6}$ | $9.88^{+1.68}_{-1.42}$ | 3.87±0.36 |
| $4^b$ | 2.82±0.09 | $1.47^{+0.09}_{-0.10}$ | 1.34±0.11 | - | - | - |
| $24^a$ | 2.89±0.11 | $1.34^{+0.09}_{-0.11}$ | 1.48±0.12 | - | - | - |

**Table 2.** Best-fit parameters for the noise components in power spectra: 1st column gives the observation numbers (same as those given in table 1). 2nd column gives the rank numbers ($S_z$) in increasing order (therefore not following the time-arrow); 3rd and 4th columns give the index and percentage rms of the power law fit (in the frequency range $0.125 - 1.0$ Hz) of VLFN respectively and the last three columns gives the cen-

| Energy (keV) | %rms | | | |
|---|---|---|---|---|
| | $S_z = 1.95 \pm 0.01$ | $S_z = 1.84 \pm 0.01$ | $S_z = 2.18 \pm 0.09$ | $S_z = 2.16 \pm 0.10$ |
| 2-5.1 | 1.99±0.19 | 1.89±0.21 | 2.29±0.12 | 1.84±0.22 |
| 5.1-7 | 3.39±0.29 | 3.57±0.34 | 3.60±0.15 | 3.31±0.27 |
| 7-10 | 4.56±0.23 | 4.32±0.38 | 4.13±0.17 | 4.21±0.25 |
| 10-16 | 3.63±0.87 | 3.75±0.6 | 5.01±0.24 | 2.98±0.95 |

**Table 3.** The photon energy dependence of %rms of PN for four different $S_z$ values.